\documentclass[aps,twocolumn,pra]{revtex4-1}

\usepackage{newlfont}
\usepackage{color}         
\usepackage{amssymb}
\usepackage{amsfonts}
\usepackage{amsmath}
\usepackage{wasysym}
\usepackage{graphicx}
\usepackage{epsfig}
\usepackage{amsthm}
\usepackage{bm}
\usepackage{mathrsfs}
\usepackage{epstopdf}
\usepackage[colorlinks=true, citecolor=blue, urlcolor=blue ]{hyperref}

\definecolor{red}{rgb}{1,0,0}
\begin{document}
\title{
Distribution of Bell inequality violation vs. multiparty quantum correlation measures
}

\author{Kunal Sharma\(^{1,2}\) Tamoghna Das\(^{2}\),  Aditi Sen(De)\(^{2}\), and Ujjwal Sen\(^{2}\)}

\affiliation{\(^1\)Indian Institute of Science Education and Research Bhopal, Bhauri, Bhopal 462 066, India}
\affiliation{\(^2\)Harish-Chandra Research Institute, Chhatnag Road, Jhunsi, Allahabad 211 019, India}


\begin{abstract}
Violation of a Bell inequality guarantees the existence of quantum correlations in a quantum state.
 A pure bipartite quantum state, having nonvanishing quantum correlation, always violates a Bell inequality. Such correspondence is absent for multipartite pure quantum states. 
For a shared multipartite quantum state, we establish a connection between the monogamy  of Bell inequality violation and genuine multi-site entanglement  as well as monogamy-based multiparty quantum correlation measures. 
We find that generalized Greenberger-Horne-Zeilinger states and another  single-parameter family  states which we refer to as the ``special Greenberger-Horne-Zeilinger'' states have  the status of extremal states in such relations.

\end{abstract}

\maketitle
\section{Introduction}

Over the last couple of decades, quantum entanglement \cite{HHHHRMP} was shown to be an useful resource  due to its vast applicability in   quantum computational  \cite{onewayQC,topologicalQC}, and  communicational tasks  \cite{densecode, teleport, crypto, amaderreview} as well as in other information processing protocols \cite{diqc, communication, Quantum_error}. 
In a large number of cases, entangled states turn out to be more advantageous in  performing  the job than the states without entanglement. 
On the other hand,  Bell had constructed a  mathematical inequality derived from  locality and reality assumptions, and showed it to be  violated by certain entangled quantum mechanical states \cite{Bell,CHSH}.

Violations of Bell-like inequalities \cite{Brunner_RMP} form necessary and sufficient criteria to detect entanglement in  pure bipartite  states \cite{Gisin}. 
The case of bipartite mixed states is more involved, and, for example, Werner states, \cite{Werner} for certain parameter ranges, do not violate the Clauser-Horne-Shimony-Holt
Bell inequality \cite{CHSH}.
 Similarly, in  multipartite systems, there are examples of  pure entangled states which do not violate multipartite correlation function \cite{WWWZB} Bell inequalities with two measurement settings  at each site \cite{Pure_multi_exce}. For multiparty quantum states, entanglement as well as Bell inequality violation of bipartite reduced states are constrained by certain monogamy relations \cite{ent_monogamy, CKW, Bell_monogamy1, Kaz_comp}. It is therefore natural to ask whether these concepts are inter-related.

In this paper, we address this query,  and in
particular, we establish a quantitative relation between a measure of nonlocal correlations quantified by the monogamy relation for the violation of Bell inequalities, called
 Bell inequality violation monogamy score (BVM)  with several  quantum correlation measures  like the 3-tangle \cite{CKW}, and quantum discord as well as quantum work deficit monogamy scores \cite{discomono, discscore, WDscore}.
Interestingly, note that Bell inequality violation can be seen as a signature of  quantum correlation stronger than entanglement while quantum discord \cite{discord} and work deficit \cite{workdeficit}  quantify weaker versions of quantum correlations beyond entanglement \cite{nonlocality_withoutent, KM_RMP}. We also establish relations between BVM score and a genuine multiparty entanglement measure quantified by the generalized geometric measure (GGM) \cite{GGM} (cf. \cite{geom_ent}). 
 For 3-qubit pure states, we identify the generalized Greenberger-Horne-Zeilinger (gGHZ) states and a one-parameter family of 
 gGHZ-like states, which we refer to  as  ``special GHZ'' states, for which the BVM scores attain minima, in different scenarios, 
  among arbitrary pure tripartite states having  the same amount of  multiparty quantum correlations. 
 Similar connections hold also  for 4-qubit pure states.
  We prove analytically that among all N-qubit symmetric states having the same GGM, the  gGHZ state possess the lowest BVM.

The paper is organised in the following way. In Sec. \ref{Sec:BVP_def}, we briefly discuss  about the Clauser-Horne-Shimony-Holt (CHSH) inequality and then define the BVM in Sec.  \ref{Sec:Monogamy_def}. In Sec. \ref{Sec:Relations}, we  establish the relations between various quantum correlation measures and BVM score. 
 In particular, Sec. \ref{Sec:BVM_GGM} deals with the connection between BVM score and genuine multiparty entanglement measures while in Sec. \ref{Sec:monogmy_monogamy}, we connect BVM score with monogamy-based measures. Finally, we conclude in Sec. \ref{Sec:conclusion}. 

\section{Bell violation parameter and Bell Monogamy score}\label{Sec:BVP_def}

Based on locality and reality assumptions, one can derive mathematical relations,   the CHSH-Bell  inequalities \cite{Bell, CHSH}, which can be shown to be violated by several quantum mechanical bipartite states. 
A bipartite state,  $\rho_{12}$, of two spin-$\frac{1}{2}$ particles ,for which a local hidden variable model exists, 
can be shown to satisfy 
\begin{equation}\label{Eq:CHSH_Ineq}
|\langle B_{CHSH}\rangle_{\rho_{12}}| \leq 2,
\end{equation}
where $B_{CHSH}=\hat{a}\,.\,\vec{\sigma}\otimes (\hat{b}+\hat{b'})\,.\,\vec{\sigma}+\hat{a'}\,.\,\vec{\sigma}\otimes(\hat{b}-\hat{b'})\,.\,\vec{\sigma}$ with four dichotomic observables, represented by four arbitrary directions, $\hat{a}, \hat{a'}$  and $\hat{b}, \hat{b'}$ and with the measurements for the observables corresponding to  $\hat{a}, \hat{a'}$ being performed by observer 1 and the remaining by observer 2. 
Here $\vec{\sigma} = (\sigma_x, \sigma_y,\sigma_z)$, with the $\sigma_\alpha$ being Pauli spin matrices, and $\langle B_{CHSH}\rangle_{\rho_{12}} = \text{Tr}(\rho_{12} B_{CHSH})$.
 
 The Hilbert-Schmidt decomposition of a two-qubit  quantum state $\rho_{12}$ is given by
\begin{eqnarray}
\rho_{12}=\frac{1}{4}\bigg( I^1 \otimes I^2+\sum_i u_i \sigma_i^1\otimes I^2+I^1\otimes \sum_i v_i \sigma_i^2 \nonumber \\
+ \sum_{n,m=1}^{3} t_{nm} \sigma^1_n\otimes \sigma^2_m\bigg), 
\end{eqnarray}
where $I$ is an identity operator, 
$t_{nm} = \text{Tr}\left( \rho_{12} \sigma^1_n \otimes \sigma^2_m\right)$ are classical correlators and $u_i$ and $v_i$ are the magnetizations. Let us denote 
the $3\times 3$ matrix formed by $t_{nm}$ as ${\cal T}_{\rho}$, and call it the correlation matrix .

For a two-qubit state $\rho_{12}$, the  maximum of $|\langle B_{CHSH}\rangle_{\rho_{12}}|$ was found to be  $2\sqrt{M(\rho_{12})}$, where $M(\rho_{12})$ is sum of the two largest eigenvalues of the symmetric matrix ${\cal T}^{\text{T}}_{\rho}{\cal T}_{\rho}$ \cite{HorodeckiPLA}.
Therefore, the quantum state violates local realism, when  $M(\rho_{12}) > 1$. For example, $M(|\psi^-\rangle) = 2$, with $|\psi^-\rangle = \frac{1}{\sqrt{2}}(|00\rangle - |11\rangle)$, giving the maximum violation.
 
For our purposes, let us consider a quantity, the Bell inequality violation parameter $(BV_{12})$, for a two-qubit state $\rho_{12}$, given by
\begin{equation}\label{Eq:BV_def}
BV_{12} = \max[2 \sqrt{M(\rho_{12})}-2, 0].
\end{equation}   
If the state $\rho_{12}$ violates a CHSH-Bell inequality, then $BV_{12}$ is nonvanishing, and  otherwise it vanishes. The above quantity will help us to write 
the Bell monogamy relation for a multiparty state which will be discussed in the  subsection below. 


\subsection{\textbf{Bell Violation Monogamy Score}}\label{Sec:Monogamy_def}
For an arbitrary N-party quantum state, $\rho_{12\ldots N}$, the monogamy score \cite{ discscore} of any bipartite measure, ${\cal M}_{12} \equiv {\cal M}(\rho_{12})$, is defined as 
\begin{equation}
\label{Eq:Monogamy_def}
\delta_{\cal M} = {\cal M}_{1:\text{rest}} - \sum_{i = 2}^N {\cal M}_{1:{i}},
\end{equation} 
 where ${\cal M}_{1:\text{rest}}$ and ${\cal M}_{1:i}$ denote the bipartite measure in the $1:\text{rest}$ bipartition and the same for two party reduced density matrices, $\rho_{1i}, ~i = 2,\ldots,N$, of the multiparty state $\rho_{12\ldots N}$.
Here the quantity, $\delta_{\cal M}$, quantifies the distribution of a given bipartite measure in a multiparty system with respect to party $1$, and we call $1$ as the nodal observer. 
 One can also define monogamy score by considering any other party as the nodal one.
The measure ${\cal M}$ is monogamous if $\delta_{\cal M} > 0$, for  arbitrary states, and otherwise it is nonmonogamous.
While certain bipartite  measures  are monogamous, there are  others that are not \cite{CKW,ent_monogamy, discomono, discscore,WDscore}.
We  now write the monogamy score for Bell inequality violation i.e., 
we replace ${\cal M}$ by $BV$. For an arbitrary  N-qubit  state $\rho$, it reads 
\begin{equation}\label{Eq:BV_monogamy_def}
\delta_{BV} =  BV_{1:\text{rest}} - \sum_{i = 2}^N BV_{1:i} \,.
\end{equation}
In this paper, we  use $\delta_{BV}$ as a quantification of the nonlocal nature present in  multiparty states.
Importantly, one should stress here that the choice of a such monogamy-based measure, results in a readily computable measure 
 for arbitrary multiqubit states, which is in general not easy for  multipartite Bell inequalities \cite{WWWZB}.

\section{Relation Between Bell inequality Violation monogamy and Several Quantum Correlation Measures}\label{Sec:Relations}

In this section, we are going to establish relations between the Bell inequality violation monogamy score and various multiparty quantum correlation measures. 
We choose two types of multiparty quantum correlation measures -- a distance-based measure, generalized geometric measure (GGM),  and several monogamy-based measures 
of quantum correlations. The GGM $(\cal E)$, a genuine multiparty entanglement measure, is defined as the minimum distance of the given state from a non-genuinely multiparty entangled state \cite{GGM}. We consider the concurrence squared monogamy score, known as 3-tangle \cite{CKW}, and quantum discord and quantum work deficit monogamy scores \cite{discscore,WDscore} as monogamy-based measures. It is known that although 3-tangle is always monogamous \cite{CKW}, quantum discord and quantum work deficit monogamy scores can be both non-negative and  negative \cite{discomono}.


\subsection{ Bell inequality Violation Monogamy with GGM}\label{Sec:BVM_GGM}
Let us begin by establishing the connection between BVM score and GGM for arbitrary N-qubit pure states. 
As we have already discussed, there exists a class of genuinely multiparty entangled states with ${\cal E} \neq 0$ which does not violate two-setting correlation function multipartite Bell 
inequalities \cite{Pure_multi_exce}.
This may lead one to believe that there is no relation between Bell inequality and genuinely multiparty entanglement in a multiparty pure state  regime. We will however  establish a universal relation between the monogamy-based Bell inequality violation score and $\cal E$, for arbitrary 3-qubit pure states.  
In particular, we find that for a fixed GGM,  $\delta_{BV}$ can not take an arbitrary value  -- it has a GGM-dependent lower limit, and thus there exists an inaccessible region in the GGM-Bell inequality monogamy score plane.    

In this investigation,  there exists two one-parameter families of N-qubit quantum states which play  important roles in determining the relevent  boundaries, given by 
\begin{equation}
\label{Eq:gGHZ_def}
|gGHZ\rangle_N = \alpha |00\ldots0\rangle_N + \sqrt{1-\alpha^2} e^{i\phi}|11\ldots 1\rangle_N,
\end{equation}  
known as generalized Greenberger-Horne-Zeilinger state \cite{GHZ},  and  
\begin{eqnarray}\label{Eq:sGHZ_def}
|sGHZ\rangle_N &=& \frac{1}{\sqrt{2}}\Big(|00\ldots0\rangle_N +  \nonumber \\
 &&  \hspace{-0.2in} |11\rangle \otimes \big(\beta |00\ldots 0\rangle + \sqrt{1-\beta^2}e^{i\theta}|11\ldots 1\rangle \big)_{N-2}\Big),\nonumber\\
\end{eqnarray}
which we call the \textit{special GHZ}   state. Here, $\alpha, \beta \in [0,1]$, and $\phi, \theta$ are phases. Before presenting the results for 3-qubit pure states, 
let us first consider the N-qubit symmetric pure states, $|\psi^{sym}\rangle_N$, 
 and  we have the following theorem.

\noindent\textbf{Theorem 1}:
\textit{If GGM of an arbitrary symmetric N-qubit pure state is equal to the GGM of the generalized GHZ state, then the Bell inequality violation monogamy score of an arbitrary symmetric state is higher or equal to that of the gGHZ state, i.e., for arbitrary N, }
\begin{equation}
\delta_{BV}(|\psi^{sym}\rangle_N) \geq \delta_{BV}(|gGHZ\rangle_N),
\end{equation}
whenever ${\cal E}(|\psi^{sym}\rangle_N) = {\cal E}(|gGHZ\rangle_N)$, for a symmetric N-qubit pure state $|\psi^{sym}\rangle_N$. 

\texttt{Proof:} 
The GGM of arbitrary symmetric state and $|gGHZ\rangle_N$  are respectively given by
\begin{equation}\label{Eq:GGM_Nsymmetric}
{\cal E}(|\psi_N^{sym}\rangle) = 1 - \max[\{e_m\}] \,\, \text{and}
\end{equation}
\begin{equation}\label{Eq:GGM_sGHZ}
{\cal E}(|gGHZ\rangle_N) = 1-\alpha,
\end{equation}
where the set, $\{e_m\}$, represents all the maximum eigenvalues of the non-repetitive marginal density matrices of the state $|\psi^{sym}\rangle_N$, and without loss of generality, we assume $\alpha \geq \frac{1}{2}$. Equating the GGMs of two states, 
we get
$\alpha = \max[\{e_m\}]$.  

We now move to calculate $\delta_{BV}$.
We write the Schmidt decomposition of  $|\psi^{sym}\rangle_N$, in the ${1:\text{rest}}$ bipartition, which is given by
\begin{equation}
|\psi^{sym}\rangle_N^{1:\text{rest}} = \sqrt{\lambda_1}|\chi\rangle_1|\xi\rangle_{N-1} +
\sqrt{1-\lambda_1}|\chi^{\perp}\rangle_1|\xi^{\perp}\rangle_{N-1},
\end{equation}
 where $\lambda_1$ is the Schmidt coefficient in the ${1:\text{rest}}$ bipartition, and we also assume $\lambda_1 \geq \frac{1}{2}$.
 It immediately gives
\begin{equation}
\label{Eq:BV_symm_N}
BV_{1:\text{rest}}(|\psi^{sym}\rangle_N) = 2\sqrt{1+4\lambda_1(1-\lambda_1)} - 2.
 \end{equation}
 It was shown \cite{Kaz_comp} that 
 among all \(\{BV_{1:i}\}_{i=2}^{N}\), at most one can be non-zero.
 Since we deal with symmetric states, all two party
 reduced density matrices are the same and hence can not violate any two settings Bell inequality.    Therefore, in this case, $\delta_{BV}(|\psi^{sym}\rangle_N)$ reduces to $BV_{1:\text{rest}}(|\psi^{sym}\rangle_N)$.
 For the $|gGHZ\rangle_N$ state, we have
\begin{equation}
\label{Eq:BV_gGHZ}
\delta_{BV}(|gGHZ\rangle_N) = 2\sqrt{1+4\alpha(1-\alpha)} - 2.
\end{equation}
 
  To obtain the relation between Eqs. (\ref{Eq:BV_symm_N}) and (\ref{Eq:BV_gGHZ}), we consider two cases.
  
\noindent\textbf{Case 1:}
Suppose that the GGM of $|\psi^{sym}\rangle_N$ comes from the single-site density matrix. So, we have $\lambda_1 = \max[\{e_m\}] = \alpha$, which leads us to $\delta_{BV}(|\psi^{sym}\rangle_N) = \delta_{BV}(|gGHZ\rangle_N)$.


\noindent\textbf{Case 2:}
If the maximum eigenvalue in the GGM comes from a reduced density matrix that corresponds to more than a  single site,  then we have $\lambda_1 \leq \alpha $. Now, since $\delta_{BV}(|\psi^{sym}\rangle_N)$ 
is a monotonically decreasing function of $\lambda$, we get
\begin{equation}
\delta_{BV}(|\psi^{sym}\rangle_N) \geq \delta_{BV}(|gGHZ\rangle_N).
\end{equation}
Hence the proof. \hfill $\blacksquare$

The above result on symmetric states leads  to the following corollary.

\noindent\textbf{Corollary 1:}
\textit{If the GGM of an arbitrary N-qubit state, coincides with that of the  gGHZ state, then the Bell inequality violation monogamy score for the arbitrary N-qubit state is bounded below by that of the gGHZ state provided
all the two-party reduced states with the nodal observer do not violate
CHSH inequalities.}  


It is known that multiparty states for which all two-party reduced states with certain observer satisfy the CHSH Bell inequalities, exist. 
In any monogamy relation with its score, given in Eq. (\ref{Eq:Monogamy_def}), if we find that all ${\cal M}_{1i} = 0 ~\forall~ i\,$, then it implies that the given state has no distribution of $\cal M$ between the nodal observer and the other single sites.
We refer to such states
as   ``non-distributive" states for that measure and that nodal observer. In this paper, in several occasions, we will focus on the properties of such multiqubit quantum states, whose bipartite reduced states do not violate any CHSH Bell inequalities. Examples include the $|gGHZ\rangle$ state. Let us now move to the case of 3-qubit pure states. We will now lift the assumptions on the symmetry property of the state and we have the following theorem.   

\begin{figure}[t]
\begin{center}
  \includegraphics[width=0.7\columnwidth,keepaspectratio,angle=270]{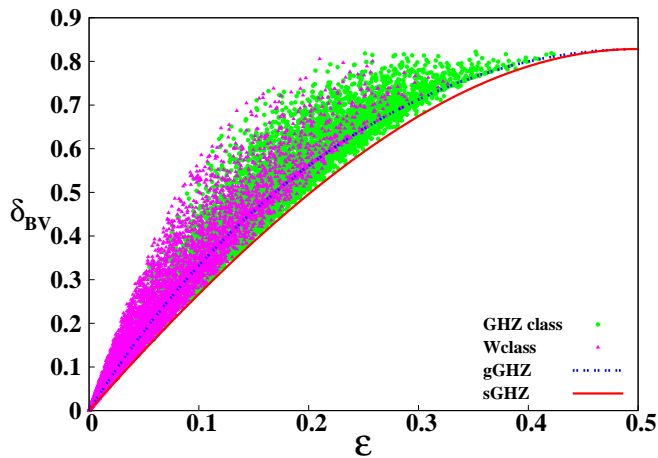}
 \end{center}
\caption{(Color.) Plot of Bell inequality violation monogamy score ($\delta_{BV}$) vs. GGM ($\cal E$) for 3-qubit pure states.  $\delta_{BV}$ is plotted along the ordinate while $\cal E$ is along the abscissa. We separately generate $10^5$ states from two distinct classes, the GHZ class (green dots) and the W class (magenta dots), and find that for both the classes of states, the special GHZ state (red solid line) gives the lower boundary. The blue dotted curve corresponds to the generalized GHZ state, which gives the boundary for those states whose maximum eigenvalue comes from the nodal part.
All axes are dimensionless.}
\label{fig:BellM_GGM_3qubit}
 \end{figure}  

\noindent \textbf{Theorem 2:} \textit{ If maximum eigenvalue required in GGM of an arbitrary 3-qubit pure state, $|\psi\rangle_{123}$, comes from the 1 : rest bipartition, then both the two-party reduced states of the tripartite state, having the observer $1$ as a party do not show any violation of the CHSH Bell inequalities}



\texttt{Proof:} Let us 
 assume that $\lambda_1 \geq \lambda_2, \lambda_3$
 where $\lambda_i $, for $i = 1,2,3$, are the maximum eigenvalues of the reduced density matrices, $\rho_i, ~i = 1,2,3$ of $|\psi\rangle_{123}$. From Eq. (\ref{Eq:BV_def}), we know that   $BV$ will be non zero only when $M(\rho_{1i}) > 1,~ i=2,3$. For any tripartite state, $|\psi\rangle_{123}$, after some algebra,  we get 
\begin{equation}\label{Eq:Mrho_relation}
M(\rho_{12}) - M(\rho_{23}) = 8\big\{\lambda_1(1-\lambda_1) - \lambda_3(1-\lambda_3)\big\}.
\end{equation} 
One can easily check that the  function $x(1-x)$ is a  monotonically decreasing function in $x \in [\frac{1}{2},1]$, and thus we have 
\begin{equation}\label{Eq:Mab_less_Mbc}
M(\rho_{12}) \leq M(\rho_{23}).
\end{equation}
But by using the monogamy of  Bell inequality violation \cite{Kaz_comp}, we know that at most one of  $\rho_{12}$ and $\rho_{23}$ can violate the CHSH Bell inequalities, and from Eq. (\ref{Eq:Mab_less_Mbc}), therefore, we get $BV_{12} = 0$. By replacing $2$ by $3$ in Eq. (\ref{Eq:Mab_less_Mbc}), one can  also show that $BV_{13} = 0$. Hence the proof. \hfill $\blacksquare$  

From Theorem 2, we can immediately establish a relation between the genuine multiparty entanglement
measure, GGM and monogamy score for CHSH Bell inequalities for 
3-qubit pure states. In particular, if the maximum eigenvalue in GGM is obtained from the
single-site density matrix of the nodal observer, i.e.,   
if for a 3-qubit  pure state, the GGM is ${\cal E}(|\psi\rangle_{123}) = 1-\lambda_1$, then from Theorem 2, we have $BV_{12} = BV_{13} = 0$. By using Eq. (\ref{Eq:BV_symm_N}), we have
$\delta_{BV}(|\psi\rangle_{123}) = 2\sqrt{1+4{\cal E}(1-{\cal E})} - 2$. 
By applying Theorems 1 and 2, we also obtain that for three-qubit pure states for which  ${\cal E}(|\psi\rangle_{123}) = 1-\lambda_1 = {\cal E}(|gGHZ\rangle_3)$, $\delta_{BV}(|\psi\rangle_{123}) = \delta_{BV}(|gGHZ\rangle_3)$ irrespective of the symmetry property of $|\psi\rangle_{123}$.

Let us now consider arbitrary  3-qubit pure states, $|\psi\rangle_{123}$, both symmetric as well as asymmetric states, and like in Theorem 1, we try to find whether there exists some lower bound on $\delta_{BV}(|\psi\rangle_{123})$ for a given value of GGM. Towards that search,
we numerically generate $2\times 10^5$ 3-qubit pure states, and calculate $\cal E$ and $\delta_{BV}$. The results are  plotted in  Fig. \ref{fig:BellM_GGM_3qubit}. Green and magenta dots represent  
 \begin{figure}[t]
\begin{center}
 \includegraphics[width=0.7\columnwidth,keepaspectratio,angle=270]{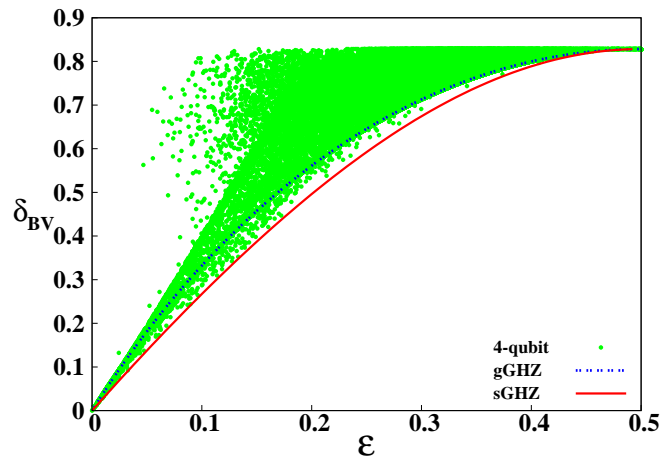}
 \end{center}
\caption{(Color online.) $\delta_{BV}$ (ordinate) against $\cal E$ (abscissa) of arbitrary 4-qubit pure states.
 We separately generate  $5\times 10^5$  4-qubit states for  all the nine classes. We find that  $|sGHZ\rangle_4$ (red solid line) gives the lower boundary. The dotted blue line represents the $|gGHZ\rangle_4$ state. All  axes are dimensionless.}
\label{fig:BellM_GGM_4qubit}
 \end{figure}      
 two SLOCC-inequivalent classes, the GHZ class and the W class \cite{zhgstate}, of 3-qubit pure states respectively, and for each class, $10^5$ states are generated  Haar uniformly.
 As shown in the figure, the entire plane of $\delta_{BV} - {\cal E}$ is not spanned by the 3-qubit pure states. 
 
 {\bf Observation:} For a fixed $\cal E $, there indeed exists  a lower boundary on $\delta_{BV}$. The lower boundary is given by $|sGHZ\rangle_3$ (red line in Fig. \ref{fig:BellM_GGM_3qubit}). In this respect,
  one should stress here that $|gGHZ\rangle_3$ (blue dotted line) does not give the lower boundary. Therefore, the symmetric states and the entire class of states possess different boundary lines of $\delta_{BV}$ for a fixed GGM.
 
We may now seek answers for two questions -- 1) does such  a lower bound exist even when one increases the number of qubits; 2) does  the one-parameter family of states which gives the lower boundary remains intact even if one changes the quantum correlation measure. We will answer the second question in the next subsection. To answer the first one, we plot in Fig. \ref{fig:BellM_GGM_4qubit}, 
 $\delta_{BV}$ against GGM when  4-qubit pure states are generated Haar uniformly. Such simulation contains all the nine 4-qubit classes, given in Ref. \cite{4qubit_class}.
 Like 3-qubit pure states, $|sGHZ\rangle_4$ again gives the lower boundary. And hence  we are tempted to conjecture that for  \emph{arbitrary N qubit pure states, if ${\cal E}(|sGHZ\rangle_N) = {\cal E}(|\psi\rangle_N)$, then $\delta_{BV}(|\psi\rangle_N) \geq \delta_{BV}(|sGHZ\rangle_N)$}.   
   

\begin{figure}[t]
\begin{center}
 \includegraphics[width=0.7\columnwidth,keepaspectratio,angle=270]{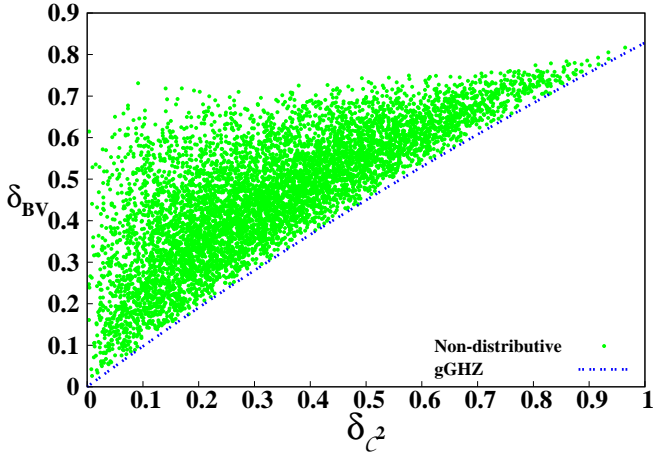}
 \end{center}
\caption{(Color online.) $\delta_{BV}$ (y-axis) vs. $\delta_{{\cal C}^2}$ (x-axis) for 3-qubit non-distributive pure states 
(green dots) with $BV_{12} = BV_{13} = 0$. 
All the points lies above the $|gGHZ\rangle_3$ state (blue dotted line). Both the axes are dimensionless. }
\label{fig:BellM_tangle_B0}
 \end{figure} 
 
\subsection{Bell inequality Violation Monogamy score with Monogamy-based Quantum Correlations}\label{Sec:monogmy_monogamy}
   
In the preceding section, we establish a connection between multiparty entanglement and Bell inequality  violation monogamy score.
 In this section, we address the question whether such feature is generic.
 Specifically, we now quantify multiparty quantum correlations by  monogamy scores, viz. 3-tangle, and quantum discord as well as quantum  work deficit monogamy scores and ask: Does $\delta_{BV}$ still possess any non-trivial lower bound? 

\subsubsection{\textbf{Connection between Bell inequality monogamy and  N-Tangle}}   
We will now show that the N-tangle and $\delta_{BV}$ are connected by the following Theorem, when N-qubit states possess a certain  symmetry. The N-tangle \cite{CKW} is defined as 
\begin{equation}
\delta_{{\cal C}^2} = {\cal C}^2_{1:rest} -\sum_{i=2}^{N} {\cal C}^2_{1:i},
\end{equation}
where ${\cal C}$ represents the concurrence \cite{Concurrence}, which is defined in the Appendix.  

\noindent \textbf{Theorem 3:} \textit{If the N-tangle of an arbitrary N-qubit pure state is same as  that of a $|gGHZ\rangle_N$ state, then the  Bell inequality violation monogamy score of the former is always bounded below by the same of the $|gGHZ\rangle_N$ state, provided all two party reduced states with the nodal party of the arbitrary N-qubit state satisfy the CHSH-Bell inequality. }

\begin{figure}[h]
\begin{center}
 \includegraphics[width=0.7\columnwidth,keepaspectratio,angle=270]{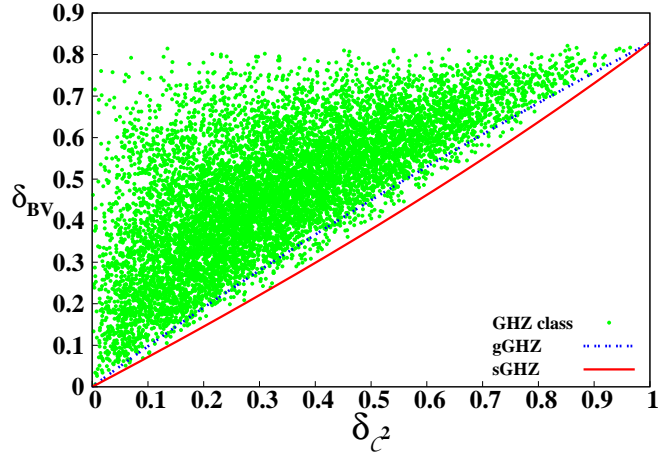}
 \end{center}
\caption{(Color online.) Plot of $\delta_{BV}$ (ordinate) with $\delta_{{\cal C}^2}$ (abcissa) for randomly generated 3-qubit GHZ class states. 
All the points lies above the $|sGHZ\rangle_3$ state (red solid line). Both  axes are dimensionless. }
\label{fig:BellM_tangle_all}
 \end{figure} 
 
\begin{figure*}[t]
 \includegraphics[width=0.7\columnwidth,keepaspectratio,angle=270]{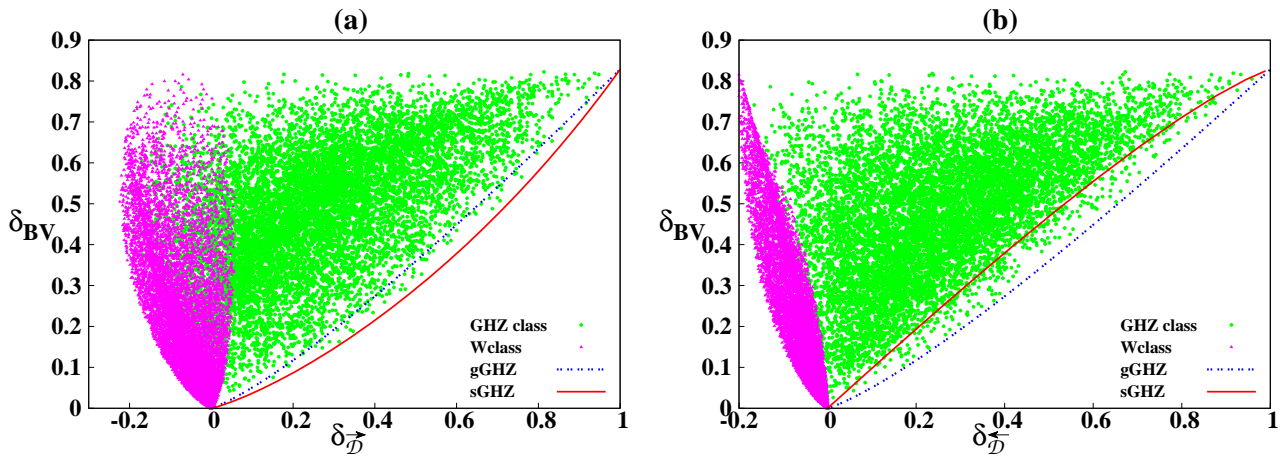}
 \caption{(Color.) Bell violation monogamy score ($\delta_{BV}$) vs. quantum discord monogamy scores $\delta_{\cal D}$ for 3-qubit pure states.  $\delta_{BV}$ is in the ordinate while $\delta_{\cal D}$ is along the abscissa. Two distinct classes of 3-qubit pure states, the GHZ class (green dots) and the W class (magenta dots) of states, are separately generated. In Fig (a), we make the measurements on the nodal part, and observe that all the points are bounded below by $|sGHZ\rangle_3$, (red solid curve), while the blue dotted line corresponds to the $|gGHZ\rangle_3$ state that lies above it. On the other hand, Fig (b), for $\delta_{\cal D}$, i.e. when the measurement in discord  are performed on the non-nodal observers,  the $|gGHZ\rangle_3$ state gives the boundary (dotted blue line). Units in abscissae are in bits while $\delta_{BV}$ is dimensionless.}
\label{fig:BellM_Discord_3qubit}
 \end{figure*}    
\texttt{Proof:} The concurrence squared monogamy score, $\delta_{{\cal C}^2}$, of  $|gGHZ\rangle_N$ state is given by $4\alpha(1-\alpha)$, while for an arbitrary  pure state, $|\psi\rangle_N$, it reads 
\begin{equation}
\delta_{{\cal C}^2}(|\psi\rangle_N) = 4\lambda_1(1-\lambda_1) - \sum_{i = 2}^N {\cal C}^2(\rho_{1:i}),
\end{equation}
where $\lambda_1$ is the maximum eigenvalue of $\rho_1 = \text{tr}_{2\ldots N}(|\psi\rangle_N\langle\psi|)$. Both the states having the same amount of $\delta_{{\cal C}^2}$ implies
\begin{eqnarray}\label{Eq:tangle_condition}
 4\lambda_1(1-\lambda_1)  &\geq & 4\alpha(1-\alpha). 
\end{eqnarray}     
Now the Bell inequality violation monogamy score of $|\psi\rangle_N$ is given by
\begin{eqnarray}
\delta_{BV}(|\psi\rangle_N) &=& 2\sqrt{1+4\lambda_1(1-\lambda_1)} - 2 \nonumber \\
                          &\geq & 2\sqrt{1+4\alpha(1-\alpha)} - 2 = \delta_{BV}(|gGHZ\rangle_N),\nonumber \\ 
\end{eqnarray}  
where we assume $\delta_{BV}(\rho_{1i}) = 0,\, i = 2,3,\ldots,N$. 
 \hfill $\blacksquare$

Theorem 3 mimics the result of symmetric states for $\delta_{{\cal C}^2}$, given in Theorem 1 and the results for 3-qubit pure states in Theorem 2. 
 Fig. \ref{fig:BellM_tangle_B0} depicts the behavior of 
$\delta_{BV}$ with  $\delta_{{\cal C}^2}$, for $10^5$ randomly generated  3-qubit non-distributive   pure states.
 The blue dotted line represents  the $|gGHZ\rangle_3$ state. 

Let us now lift the constraint of $BV(\rho_{1i}) = 0$.  
By numerically 
generating  $10^5$ tripartite pure states,  from the GHZ class (green dots) by using Haar measure,
 we observe that like for the GGM,
the $|sGHZ\rangle_3$ state has a special status, and we find  
$\delta_{BV} (|\psi\rangle_{123}) \geq  \delta_{BV} (|sGHZ\rangle_{123})$,
when $\delta_{{\cal C}^2} (|\psi\rangle_{123}) =  \delta_{{\cal C}^2} (|sGHZ\rangle_{123})$, as shown in Fig. \ref{fig:BellM_tangle_all}.  


\subsubsection{\textbf{Bell inequality Monogamy with quantum Discord and quantum Work deficit monogamy}}

Upto now, we found that the relations between Bell inequality violation monogamy and the entanglement are independent of the choice of measure of the entanglement. We will now go beyond entanglement  and find whether such relation holds or not.
Specifically, instead of entanglement, we now consider quantum discord monogamy \cite{discomono, discscore}, and quantum work deficit monogamy  scores \cite{WDscore}, denoted respectively by \(\delta_D\) and \(\delta_{WD}\). 
Note that both quantum  discord, and quantum work deficit involve local measurements on one of the subsystems of the bipartite state.  For $\rho_{12}$, 
 if for defining quantum discord, measurement is performed on party  $2$, we denote discord by $\overleftarrow{\cal D}$ and corresponding monogamy score as  
$\delta_{\overleftarrow{\cal D}}$.  $\delta_{\overrightarrow{\cal D}}$ represents the monogamy score when the measurement is on the party $1$. 
Similarly, we  have $\delta_{\overleftarrow{\cal WD}}$, and $\delta_{\overrightarrow{\cal WD}}$ .

For an arbitrary 3-qubit pure state, we find that both the information-theoretic monogamy scores follow similar relations as obtained in Theorems 2 and 3, irrespective of the party chosen for measurement. 
However, the one-parameter family of the boundary states changes, depending on the choice of the party in which measurements are performed.
In particular, when measurement has been done on the nodal part, i.e. $1$ in our case,  $|sGHZ\rangle_3$ gives the lower boundary while the $|gGHZ\rangle_3$  
gives the lower boundary when measurements have been carried out on the other   parties i.e. $2$ or $3$. These results are depicted in Figs \ref{fig:BellM_Discord_3qubit} (a) and (b).  

\textbf{Theorem 4:} \textit{The Bell inequality violation monogamy scores of arbitrary non-distributive N-qubit pure states, having the same quantum discord as well as  quantum work deficit monogamy scores, are bounded below by those of the $|gGHZ\rangle_N$ states.}

\texttt{Proof:} Quantum discord monogamy score of the N-qubit $|gGHZ\rangle$ state is given by $\delta_{\overrightarrow{\cal D}}(|gGHZ\rangle_N) = \delta_{\overleftarrow{\cal D}}(|gGHZ\rangle_N) = H(\alpha)$, 
where $H(\alpha)$ is the binary entropy having the form
\begin{equation}
H(\alpha) = -\alpha\log_2(\alpha) - (1-\alpha)\log_2(1-\alpha),
\end{equation}
for $0\leq \alpha \leq 1$. 
On the other hand, $\delta_{\overrightarrow{\cal D}}$ for an non-distributive arbitrary pure state, $|\psi\rangle_N$,  is given by
\begin{eqnarray}
\delta_{\overrightarrow{\cal D}}(|\psi\rangle_N) &=& S(\rho_1) - \sum_i{\overrightarrow{\cal D}}(\rho_{1:i}) \nonumber \\
&\leq & H(\lambda_1),
\end{eqnarray}
where we have used the fact that ${\overrightarrow{\cal D}}\geq 0$, for all the bipartite states, and $S(\rho_1) = H(\lambda_1)$ is the von-Neumann entropy of the marginal density matrix of the nodal part. Also, 
$\delta_{\overleftarrow{\cal D}}(|\psi\rangle_N) \leq H(\lambda_1) $. 
 Equating $\delta_{\overleftarrow{\cal D}}(|gGHZ\rangle_N)$ with $\delta_{\overleftarrow{\cal D}}(|\psi\rangle_N)$, we obtain $\alpha \geq \lambda_1 $, assuming $\alpha,\lambda_1 \geq \frac{1}{2}$. 
Now from Eqs.  (\ref{Eq:BV_symm_N}) and (\ref{Eq:BV_gGHZ}), we conclude 
$\delta_{BV}(|\psi\rangle_N) \geq \delta_{BV}(|gGHZ\rangle_N)$ for non-distributive states.

Now for the quantum work deficit monogamy score, one can also show that $\delta_{\overrightarrow{\cal WD}}(|\psi\rangle_N) \leq H(\lambda_1)$, as well 
as that $\delta_{\overleftarrow{\cal WD}}(|\psi\rangle_N) \leq H(\lambda_1)$, 
and using  similar argument as above, for non-distributive states, we have $\delta_{BV}(|\psi\rangle_N) \geq \delta_{BV}(|gGHZ\rangle_N)$. Hence the proof. \hfill $\blacksquare$

We again numerically generate $10^5$ arbitrary 3-qubit pure states, both from the GHZ class as well as from the W class states, and plot $\delta_{BV}$ with respect to $\delta_{\overrightarrow{\cal D}}$ and $\delta_{\overleftarrow{\cal D}}$ in  Figs. \ref{fig:BellM_Discord_3qubit} (a) and \ref{fig:BellM_Discord_3qubit} (b) respectively.
As already mentioned, the choice of party in which measurements are performed play  an important role in determining the boundary line of the scattered points.


\section{Conclusion}\label{Sec:conclusion}

In summary, we considered the relation between a monogamy-based measure of multipartite nonlocal correlations and several monogamy-based neasures of multipartite quantum correlations, for multiparty quantum states. Here ``nonlocal correlations'' have to be understood as correlations that violate local realism. We also established a connection between the multipartite nonlocal correlations and a measure of genuine multiparty entanglement. 
We found that the generalized GHZ states and a single-parameter family of states, which we call ``special GHZ'' states play  important roles in these relations.

\begin{acknowledgments}
KS thanks the Harish-Chandra Research Institute (HRI) for giving him the opportunity to visit the quantum information and computation  group. We acknowledge computations performed at the cluster computing facility at HRI. 
\end{acknowledgments}

\section*{Appendix}

We now briefly define all the multiparty quantum correlation measures that were required in this work. We first discuss about the distance-based genuine multiparty entanglement measure and then we focus on the monogamy-based quantum correlation measures.

\subsection*{Genuine Multiparty Entanglement Measure}
For an arbitrary N-party pure state, $|\psi\rangle_N$, the GGM ($\cal E$) \cite{GGM} (cf. \cite{geom_ent}) is obtained as the minimum distance from a pure state which is not genuinely multiparty entangled and reads as 
\begin{equation}\label{Eq:GGM_def}
\mathcal{E}(\vert\psi\rangle_N)= \min_{|\chi\rangle \in {\cal S}}\big[1-|\langle\chi\vert\psi_N\rangle|^2\big]
\end{equation}
Here the minimization is taken over the set, ${\cal S}$, of all non-genuinely multiparty pure states. 
Eq. (\ref{Eq:GGM_def}) reduces to a simplified form which makes the measure computable for arbitrary number of parties and is given by 
\begin{equation}
\mathcal{E}(\vert\psi\rangle_N) = 1-\max\lbrace{\lambda_{A:B}\vert A\cup B= \lbrace{1, 2,.., N}\rbrace, A\cap B=\emptyset}\rbrace,
\end{equation}
where $\lambda_{A:B}$ is the maximum eigenvalue of the marginal density matrices  of $|\psi\rangle_N$.

\subsection*{Monogamy-based Quantum Correlations}
Let us now discuss about the monogamy-based quantum correlation measures of an N-partite quantum states. In Eq. (\ref{Eq:Monogamy_def}) of Sec. \ref{Sec:Monogamy_def}, we introduced the monogamy scores $\delta_{\cal M}$, for an arbitrary quantum correlation measure $\cal M$. 
We used the  monogamy scores for concurrence squared, quantum discord and quantum work deficit. We will now briefly define these bipartite measure.


\subsubsection*{\textbf{Concurrence }}
Concurrence is a bipartite entanglement measure introduced by Bennett et. al.  \cite{Concurrence}, for $2\otimes d$ system. It is a monotonic function of entanglement of formation and an entanglement monotone under local operations and classical communication. 
For an arbitrary 2-qubit mixed state, $\rho_{12}$, concurrence ($\cal C $)  is defined as
\begin{equation}
{\cal C}(\rho_{12}) = \max \big\{0, \lambda_1 - \lambda_2 - \lambda_3-\lambda_4\big\}
\end{equation}  
where $\lambda_i$ are the square root of the eigenvalues of the non-Hermitian operator $\rho_{12}\tilde{\rho}_{12}$, in descending order, and $\tilde{\rho}_{12} = (\sigma_y \otimes \sigma_y) \rho_{12}^* (\sigma_y \otimes \sigma_y)$ is the spin flipped state. This expression can be simplified for arbitrary 2-qubit pure states $|\psi_{12}\rangle$, and is given by $2\sqrt{\text{det}(\rho_1)}$, where $\rho_1 = \text{tr}_2(|\psi_{12}\rangle\langle\psi_{12}|) $ \cite{ent_monogamy, CKW}.

The N-tangle or the concurrence squared monogamy score is obtained from Eq. (\ref{Eq:Monogamy_def}), by replacing $\cal M$ by ${\cal C}^2$ \cite{Concurrence}. For 3-qubit pure states, the 3 tangle vanishes for the W class states. 


\subsection*{\textbf{Quantum Discord}}
Quantum discord of an arbitrary bipartite quantum state, $\rho_{12}$, is defined as the difference between the total correlation and the classical correlation present in the state.
For $\rho_{12}$, the total correlation is quantified  as the minimum amount  of 
noise required  to make $\rho_{12}$  into a product state of the form $\rho_1 \otimes \rho_2$ \cite{totalcorre}, where $\rho_1$ and $\rho_2$, are the marginal density matrices. 
\begin{equation}
\textit{I}(\rho_{12})=S(\rho_1)+S(\rho_2)-S(\rho_{12}),
\end{equation}
where $S(\rho) = -\text{tr}(\rho\log_2 \rho)$ is the von-Neumann entropy of $\rho$. $\textit{I}(\rho_{12})$ is also known as the quantum mutual information of $\rho_{12}$. In a similar spirit, the classical correlation is  quantified as the amount of noise that has to be introduced to make $\rho_{12}$ classically correlated \cite{totalcorre}, and is given by
\begin{equation}
J(\rho_{12})=S(\rho_1)-S(\rho_{1|2}),
\end{equation}
where $S(\rho_{1|2})$ is the conditional entropy, defined as
$ S(\rho_{1|2})=\min_{\lbrace\prod_i^2\rbrace} \sum{p_i S(\rho_{1|i})}$, with the minimum being taken over all rank one projectors $\{\prod_i^2\}$, acting on the subsystem $2$, and $\rho_{1|i} =\frac{1}{p_i} \Big(I^1\otimes \prod_i^2 \rho_{12}  I^1\otimes \prod_i^2\Big)$. Here $p_i = \text{tr}\big(I^1\otimes \prod_i^2 \rho_{12}  I^1\otimes \prod_i^2\big)$. Quantum discord is then defined as \cite{discord}
\begin{equation}
\overleftarrow{\cal D}(\rho_{12}) = I(\rho_{12}) - J(\rho_{12}).
\end{equation} 
 The left arrow in ${\cal D}$ indicates that the measurement is performed  on the second party $2$. Similarly,  we can have $\overrightarrow{\cal D}$, when  measurement occurs on the first party. 

\subsection*{\textbf{Quantum Work deficit}}
Like quantum discord, quantum work deficit of a bipartite state $\rho_{12}$, is  the difference between two quantities, the extractable work from a quantum state under  ``closed operations" (CO) and  ``closed local operations and classical communication" (CLOCC) \cite{workdeficit}. 
The operations in CO include (i) global unitary operations,
and (ii) dephasing of $\rho_{12}$ by the projective operators defined in the Hilbert space of $\rho_{12}$, while CLOCC involves (i) local unitary operations, (ii) dephasing by local measurements on the subsystem $1$ or $2$, and (iii) communicating the dephased subsystem to the complementary subsystem $2$ or $1$, over a noiseless quantum channel.
Under CO, the amount of extractable work from $\rho_{12}$, is 
\begin{equation}
I_{CO}(\rho_{12}) = \log(\mbox{dim}{\cal H}) - S(\rho_{12}),
\end{equation} 
with $\mbox{dim}{\cal H}$ being the dimension of the Hilbert space of $\rho_{12}$, while under CLOCC, it is given by
\begin{equation}
I_{CLOCC}(\rho_{12}) = \log(\mbox{dim}{\cal H}) - \min_{\prod^B_i}S(\rho^{'}_{12})
\end{equation} 
where $\rho^{'}_{12} = \sum_i I^1\otimes\prod^2_i \rho_{12} I^1\otimes\prod^2_i$
Now the quantum work deficit is given by
\begin{equation}
\overleftarrow{\cal WD}(\rho_{12}) = I_{CO}(\rho_{12}) - I_{CLOCC}(\rho_{12}).
\end{equation}
Similarly one can have $\overrightarrow{\cal WD}(\rho_{12})$, by changing the subsystem at which the measurement is performed.


\begin{thebibliography}{19}
\bibitem{HHHHRMP} R. Horodecki, P. Horodecki, M. Horodecki, and K. Horodecki,
Rev. Mod. Phys. {\bf 81}, 865 (2009).


 
\bibitem{onewayQC} R. Raussendorf and H. J. Briegel, Phys. Rev. Lett. {\bf 86}, 5188 (2001); 
                  P. Walther, K. J. Resch, T. Rudolph, E. Schenck, H. Weinfurter, V. Vedral, M. Aspelmeyer, and A. Zeilinger, Nature {\bf 434}, 169 (2005); 
                  H. J. Briegel, D. Browne, W. D\"ur, R. Raussendorf,  and M. van den Nest, Nat. Phys. {\bf 5}, 19 (2009).
                  
\bibitem{topologicalQC} A. Kitaev, Ann. Phys. (NY) {\bf 303}, 2 (2003). 
 

\bibitem{densecode}C. H. Bennett and S. J. Wiesner, Phys. Rev. Lett. \textbf{69}, 2881 (1992);
                   K. Mattle, H. Weinfurter, P. G. Kwiat, and A. Zeilinger, Phys. Rev. Lett. \textbf{76}, 4656 (1996).



\bibitem{teleport}C. H. Bennett, G. Brassard, C. Cr\'{e}peau, R. Jozsa, A. Peres, and W. K. Wootters, Phys. Rev. Lett. \textbf{70}, 1895 (1993);
                  D. Bouwmeester, J. W. Pan, K. Mattle, M. Eibl, H. Weinfurter, and A. Zeilinger, Nature \textbf{390}, 575 (1997);
                  J. W. Pan, D. Bouwmeester, H. Weinfurter, and A. Zeilinger, Phys. Rev. Lett. \textbf{80}, 3891 (1998);
                  D. Bouwmeester, J. W. Pan, H. Weinfurter, and A. Zeilinger, J. Mod. Opt. \textbf{47}, 279 (2000).                       
                       

\bibitem{crypto}A. Ekert, Phys. Rev. Lett. \textbf{67}, 661 (1991);
                T. Jennewein, C. Simon, G. Weihs, H. Weinfurter, and A. Zeilinger, Phys. Rev. Lett. \textbf{84}, 4729 (2000);
                D. S. Naik, C. G. Peterson, A. G. White, A. J. Berglund, and P. G. Kwiat, Phys. Rev. Lett. \textbf{84}, 4733 (2000);
                W. Tittel, T. Brendel, H. Zbinden, and N. Gisin, Phys. Rev. Lett. \textbf{84}, 4737 (2000);
                N. Gisin, G. Ribordy, W. Tittel, and H. Zbinden, Rev. Mod. Phys. {\bf 74}, 145 (2002).

\bibitem{amaderreview} For a recent review on quantum communication, see e.g. A. Sen(De) and U. Sen, Physics News {\bf 40}, 17 (2010) (arXiv:1105.2412).


\bibitem{diqc}A. Ac\'{\i}n, N. Brunner, N. Gisin, S. Massar, S. Pironio, and V. Scarani, Phys. Rev. Lett. {\bf 98}, 230501 (2007);
              S. Pironio, A. Ac\'{\i}n, N. Brunner, N. Gisin, S. Massar, and V. Scarani, New J. Phys. \textbf{11}, 045021 (2009);
              N. Gisin, S. Pironio, and N. Sangouard, Phys. Rev. Lett. {\bf 105}, 070501 (2010);
              L. Masanes, S. Pironio, and A. Ac\'{\i}n, Nature Commun. \textbf{2}, 238 (2011);
              H. K. Lo, M. Curty, and B. Qi, Phys. Rev. Lett. {\bf 108}, 130503 (2012).


\bibitem{communication}M. {\.Z}ukowski,  A. Zeilinger, M. Horne, and H. Weinfurter, Acta Phys. Pol. {\bf 93}, 187 (1998); 
M. Hillery, V. Buzek, and A. Berthiaume, Phys. Rev. A {\bf 59}, 1829 (1999); 
R. Demkowicz-Dobrzanski, A. Sen(De), U. Sen, and M. Lewenstein,  \emph{ibid.} {\bf 80}, 012311 (2009); 
R. Cleve, D. Gottesman, and H.-K. Lo, Phys. Rev. Lett. {\bf 83}, 648 (1999); 
A. Karlsson, M. Koashi, and N. Imoto, Phys. Rev. A {\bf  59}, 162 (1999).


\bibitem{Quantum_error}  A. R. Calderbank and P.W. Shor, Phys. Rev. A \textbf{54}, 1098 (1996);
A. M. Steane, Phys. Rev. Lett. \textbf{77}, 793 (1996); 
A. M. Steane, Phys. Rev. A  \textbf{54}, 4741 (1996);
A. R. Calderbank, E.M. Rains, P. W. Shor, and N. J. A. Sloane, Phys. Rev. Lett. {\bf 78}, 405 (1997). 


\bibitem{Bell}J. S. Bell, Physics {\bf 1}, 195 (1965).

\bibitem{CHSH} J. F. Clauser, M. A. Horne, A. Shimony, and R. A. Holt
Phys. Rev. Lett. {\bf 23}, 880 (1969).



\bibitem{Brunner_RMP} N. Brunner, D. Cavalcanti, S. Pironio, V. Scarani, and S. Wehner, Rev. Mod. Phys {\bf 86}, 419 (2014). 

\bibitem{Gisin} N. Gisin Phys. Lett. A {\bf 154}, 201 (1991); N. Gisin, A. Peres, Phys. Lett. A {\bf 162}, 15 (1992).


\bibitem{Werner} R. F. Werner, Phys. Rev. A {\bf 40}, 4277 (1989).

\bibitem{WWWZB} H. Weinfurter and M. {\.Z}ukowski, Phys. Rev. A {\bf 64}, 010102
(2001); R. F. Werner and M. M. Wolf, \emph{ibid}. {\bf 64}, 032112 (2001);
M. {\.Z}ukowski and C. Brukner, Phys. Rev. Lett. {\bf 88}, 210401 (2002).


\bibitem{Pure_multi_exce} M. {\.Z}ukowski, {\^C}. Brukner, W. Laskowski, and M. Wie{\'s}niak, Phys. Rev. Lett. {\bf 88}, 210402 (2002); A. Sen(De), U. Sen, and M. {\.Z}ukowski, Phys. Rev. A 66, 062318 (2002).



\bibitem{ent_monogamy} C. H. Bennett, H. J. Bernstein, S. Popescu, and B. Schu- macher, Phys. Rev. A \textbf{53}, 2046 (1996); 
M. Koashi and A. Winter, Phys. Rev. A \textbf{69}, 022309 (2004);
Y. -C. Ou and H. Fan, Phys. Rev. A \textbf{75}, 062308 (2007);
 G. Adesso, A. Serafini, and F. Illuminati, Phys. Rev. A \textbf{73}, 032345 (2006); T. Hiroshima, G. Adesso, and F. Illuminati, Phys. Rev. Lett. \textbf{98}, 050503 (2007); ; M. Hayashi and L. Chen, Phys. Rev. A \textbf{84}, 012325 (2011);  A. Streltsov, G. Adesso, M. Piani, and D Bru{\ss}, Phys.
Rev. Lett. \textbf{109}, 050503 (2012); F. F. Fanchini, M. C. de Oliveira, L. K. Castelano, and M. F. Cornelio, Phys. Rev. A \textbf{87}, 032317 (2013);
 Y.-K. Bai, Y.-F. Xu, and Z. D. Wang
Phys. Rev. Lett. \textbf{113}, 100503 (2014); B. Regula, S. D. Martino, S. Lee, and G. Adesso, Phys. Rev. Lett. \textbf{113}, 110501 (2014).

\bibitem{CKW} V. Coffman, J. Kundu, and W. K. Wootters, Phys. Rev. A {\bf 61}, 052306 (2000);  T. Osborne and F. Verstraete, Phys. Rev. Lett. \textbf{96}, 220503 (2006).





\bibitem{Bell_monogamy1} B. Toner and F. Verstraete, quant-ph/0611001;
J. Barrett, N. Linden, S. Massar, S. Pironio, S. Popescu, and D. Roberts, Phys. Rev. A \textbf{71}, 022101 (2005);
Ll. Masanes, A. Acin, and N. Gisin, Phys. Rev. A \textbf{73}, 012112 (2006);
M. Seevinck, Phys. Rev. A \textbf{76}, 012106 (2007); S. Lee and J. Park, Phys. Rev. A \textbf{79}, 054309 (2009); A. Kay, D. Kaszlikowski, and R. Ramanathan, Phys. Rev. Lett. \textbf{103}, 050501 (2009);
B. Toner, Proc. R. Soc. A \textbf{465}, 59 (2009);
M. Pawlowski and {\^C}. Brukner, Phys. Rev. Lett. \textbf{102}, 030403 (2009).

\bibitem{Kaz_comp} P. Kurzyński, T. Paterek, R. Ramanathan, W. Laskowski, and D. Kaszlikowski, Phys. Rev. Lett. {\bf 106}, 180402 (2011).








\bibitem{discomono} R. Prabhu, A. K. Pati, A. Sen(De), and U. Sen, Phys.
Rev. A \textbf{85}, 040102(R) (2012); G. L. Giorgi, Phys. Rev. A \textbf{84}, 054301 (2011);  F. F. Fanchini, M. F. Cornelio, M. C. de Oliveira, and A.
O. Caldeira, Phys. Rev. A \textbf{84}, 012313 (2011);
R. Prabhu, A. K. Pati, A. Sen(De), and U. Sen,
Phys. Rev. A \textbf{86}, 052337 (2012);
 H. C. Braga, C. C. Rulli, T. R. de Oliveira, and M. S. Sarandy,
Phys. Rev. A \textbf{86}, 062106 (2012); S.-Y. Liu, Y.-R. Zhang, L.-M. Zhao, W.-L. Yang, and H. Fan,
Ann. Phys. \textbf{348}, 256 (2014).

\bibitem{discscore} M. N. Bera, R. Prabhu, A. Sen(De), and U. Sen, Phys. Rev. A {\bf 86}, 012319 (2012).
\bibitem{WDscore} Salini K., R. Prabhu, A. Sen(De), and U. Sen, Ann. Phys. {\bf 348}, 297 (2014); A. Kumar, R. Prabhu, A. Sen(De), and U. Sen, Phys. Rev. A {\bf 91}, 012341 (2015). 

\bibitem{discord} L. Henderson and V. Vedral, J. Phys. A {\bf 34}, 6899 (2001);  
	H. Ollivier and W.H. Zurek, Phys. Rev. Lett. {\bf 88}, 017901 (2001). 
	
\bibitem{workdeficit} J. Oppenheim, M. Horodecki, P. Horodecki and R. Horodecki, Phys. Rev. Lett. {\bf 89}, 180402 (2002); M. Horodecki, K. Horodecki, P. Horodecki, R. Horodecki, J. Oppenheim, A. Sen(De), and U. Sen, Phys. Rev. Lett. {\bf 90}, 100402 (2003); I. Devetak, Phys. Rev. A {\bf 71}, 062303 (2005); M. Horodecki, P. Horodecki, R. Horodecki, J. Oppenheim, A. Sen(De), U. Sen, and B. Synak-Radtke, Phys. Rev. A {\bf 71}, 062307 (2005).




\bibitem{nonlocality_withoutent} C. H. Bennett, D. P. DiVincenzo, C. A. Fuchs, T. Mor, E. Rains, P. W. Shor, J. A. Smolin, and W. K. Wootters, Phys. Rev. A {\bf 59}, 1070 (1999); 
                  C. H. Bennett, D.P. DiVincenzo, T. Mor, P.W. Shor, J.A. Smolin, and B.M. Terhal, Phys. Rev. Lett. {\bf 82}, 5385 (1999); J. Walgate, A.J. Short, L. Hardy, and V. Vedral,ibid. \textbf{85}, 4972 (2000); 
                  S. Virmani, M.F. Sacchi, M.B. Plenio, and D. Markham, Phys. Lett. A \textbf{288}, 62 (2001); 
                  Y.-X. Chen and D. Yang, Phys. Rev. A \textbf{64}, 064303 (2001); ibid. \textbf{65}, 022320 (2002); 
                  J. Walgate and L. Hardy, Phys. Rev. Lett. \textbf{89}, 147901 (2002); D. P. DiVincenzo, T. Mor, P. W. Shor, J. A. Smolin, and B. M. Terhal Commun. Math. Phys. {\bf 238}, 379 (2003);
                  M. Horodecki, A. Sen(De), U. Sen, and K. Horodecki, ibid. \textbf{90}, 047902 (2003); 
                  W. K. Wootters, Int. J. Quantum Inf. \textbf{4}, 219 (2006).

\bibitem{KM_RMP} K. Modi, A. Brodutch, H. Cable, T. Paterek, and V. Vedral, Rev. Mod. Phys. \textbf{84}, 1655 (2012), and  references therein.              


\bibitem{GGM} A. Sen(De) and U. Sen, Phys. Rev. A \textbf{81}, 012308 (2010);
A. Sen(De) and U. Sen, 
arXiv:1002.1253 [quant-ph].

\bibitem{geom_ent} A. Shimony, Ann. NY Acad. Sci. \textbf{755}, 675 (1995); 
                   H. Barnum and N. Linden, J. Phys. A \textbf{34}, 6787 (2001); 
                   T. -C. Wei, and P. M. Goldbart, Phys. Rev. A \textbf{68}, 042307 (2003);
                   T. -C. Wei, M. Ericsson, P. M. Goldbart, and W. J. Munro, 
                   Quantum Inform. Compu. \textbf{04}, 252 (2004); 
                   D. Cavalcanti, Phys. Rev. A \textbf{73}, 044302 (2006);
                   M. Blasone, F. Dell'Anno, S. DeSiena, and F. Illuminati, Phys. Rev. A \textbf{77}, 062304 (2008).                   

\bibitem{HorodeckiPLA} R. Horodecki, P. Horodecki, M. Horodecki, Phys. Lett. A, {\bf 200}, 340 (1995).


\bibitem{GHZ} D.M. Greenberger, M.A. Horne, and A. Zeilinger, in Bell’s Theorem, Quantum Theory, and Conceptions of
the Universe, ed. M. Kafatos (Kluwer Academic, Dordrecht, The Netherlands, 1989).


\bibitem{zhgstate}A. Zeilinger, M. A. Horne, and D. M. Greenberger, 
                   in Proceedings of Squeezed States and Quantum Uncertainty, 
                   edited by D. Han, Y. S. Kim, and W. W. Zachary, NASA Conf. Publ. \textbf{3135}, 73 (1992); W. D\"{u}r, G. Vidal, J. I. Cirac, Phys. Rev. A \textbf{62}, 062314 (2000).                  

\bibitem{4qubit_class} F. Verstraete, J. Dehaene, B. De Moor, and H. Verschelde
Phys. Rev. A {\bf 65}, 052112 (2002).

\bibitem{Concurrence} C. H. Bennett, D. P. DiVincenzo, J. Smolin, and W. K. Wootters, Phys. Rev. A {\bf 54}, 3824 (1996); S. Hill and W. K. Wootters, Phys.
Rev. Lett. {\bf 78}, 5022 (1997); W. K. Wootters, Phys. Rev. Lett. {\bf 80},
2245 (1998); W. K. Wootters, Quant. Inf. Comput. {\bf 1}, 27 (2001).


%
%
%
%
%
%
%
%
%
%
%
%
%
%
%
%

%





\bibitem{totalcorre} B. Groisman, S. Popescu, and A. Winter
Phys. Rev. A {\bf 72}, 032317 (2005).

\end{thebibliography}
\end{document}